\documentclass[twocolumn,showpacs,aps,epsfig,nofootinbib]{revtex4}
\usepackage{graphicx}
\usepackage{epstopdf}
\usepackage{latexsym}
\usepackage{amssymb}
\usepackage[center]{subfigure}

\begin{document}

%%%%%%%%%%%%%%%%%%%%%%%%%%%%%%%%%%%%%%%%%%%%%%%%%%%%%%%%%%%%%%%
 \newcommand{\bq}{\begin{equation}}
 \newcommand{\eq}{\end{equation}}
 \newcommand{\bqn}{\begin{eqnarray}}
 \newcommand{\eqn}{\end{eqnarray}}
 \newcommand{\nb}{\nonumber}
 \newcommand{\lb}{\label}
\newcommand{\PRL}{Phys. Rev. Lett.}
\newcommand{\PL}{Phys. Lett.}
\newcommand{\PR}{Phys. Rev.}
\newcommand{\CQG}{Class. Quantum Grav.}
 %%%%%%%%%%%%%%%%%%%%%%%%%%%%%%%%%%%%%%%%%%%%%%%%%%%%%%%%%%%%%%%

\title{Static electromagnetic fields and charged black holes in  general covariant theory of  Ho\v{r}ava-Lifshitz gravity}

\author{Ahmad Borzou $^{a}$}
\email{ahmad_borzou@baylor.edu}

\author{Kai Lin $^{a, b}$}
\email{k_lin@baylor.edu}

\author{Anzhong Wang $^{a, c}$}
\email{anzhong_wang@baylor.edu}

\affiliation{$^{a}$ GCAP-CASPER, Physics Department, Baylor
University, Waco, TX 76798-7316, USA\\
$^{b}$ Institute of Theoretical Physics, Chongqing University, Chongqing 400030, China\\
$^{c}$ Institute  for Advanced Physics $\&$ Mathematics,
Zhejiang University of
Technology, Hangzhou 310032,  China}

\date{\today}

\begin{abstract}

In this paper, we study electromeganetic  static spacetimes   in the nonrelativisitc general covariant theory of the Ho\v{r}ava-Lifshitz (HL) gravity, proposed 
recently by Ho\v{r}ava and Melby-Thompson, and present all the  electric static solutions, which represent the generalization of the Reissner-Nordstr\"om solution 
found in Einstein's  general relativity (GR). The global/local structures of spacetimes in the HL theory in general are different from those given in GR, because
the dispersion relations of test particles now  contain high-order momentum terms, so the speeds of these particles are unbounded in the ultraviolet  (UV). As 
a result,  the conception of light-cones defined in GR becomes invalid and test particles  do not follow geodesics. To study black holes in the HL theory, we 
adopt the geometrical optical approximations, and define a horizon as   a (two-closed) surface that is free of spacetime singularities and on which massless
test particles are infinitely redshifted. With such a definition, we show that  some of our solutions give rise to (charged) black holes, although the radii of their
 horizons in general depend on the energies of the test particles. 

\end{abstract}


\maketitle

\section{Introduction}
\renewcommand{\theequation}{1.\arabic{equation}} \setcounter{equation}{0}

Recently, Ho\v{r}ava proposed a  theory of quantum gravity \cite{Horava}, motivated by the Lifshitz scalar field
theory in  solid state physics \cite{Lifshitz}. %, for which the theory is often referred to as
Due to several remarkable features, the HL theory has  attracted a great deal
of attention (see for example, \cite{reviews,Mukc} and references therein).
In this  theory,  the general covariance is   broken down to  the  foliation-preserving
diffeomorphisms Diff($M, \; {\cal{F}}$),
\bq
\lb{1.4}
\tilde{t} =  f(t),\; \;\; \tilde{x}^{i}  =  \zeta^{i}(t, {\bf x}),
\eq
because of which, in comparison with GR one more degree of freedom appears
in the gravitational sector -  the spin-0 graviton. This is potentially dangerous, and needs to decouple
in the infrared  (IR), in order to be consistent with observations.   Whether this is possible or not is still an
open question \cite{reviews}.  In particular,  Mukohyama studied the spherically symmetric static spacetimes \cite{Mukc}, and showed
explicitly
that the spin-0 graviton indeed decouples after nonlinear effects are taken into account, an analogue of the
Vainshtein effect  \cite{Vain}, initially found in massive gravity \cite{BDZ}. Similar considerations in cosmology were  presented
in \cite{Izumi:2011eh,GMW} (See also \cite{WWa} for a class of exact solutions.), where a fully nonlinear analysis of superhorizon cosmological perturbations was carried out,
by adopting the so-called gradient expansion method \cite{SBLMS}. It was found that   the relativistic  limit of the HL theory  is continuous,
and GR is recovered at least in two different cases: (a) when only  the ``dark matter as an integration constant'' is present \cite{Izumi:2011eh};
and (b) when a scalar field and    the ``dark matter as an integration constant'' are present \cite{GMW}.

Another very attractive approach is to eliminate the spin-0 graviton by introducing two auxiliary fields,
the $U(1)$ gauge field $A$ and the Newtonian prepotentail $\varphi$, by   extending
the  Diff($M, \; {\cal{F}}$) symmetry (\ref{1.4}) to include  a local $U(1)$ symmetry \cite{HMT},
\bq
\lb{symmetry}
 U(1) \ltimes {\mbox{Diff}}(M, \; {\cal{F}}).
 \eq
Under this extended symmetry,   the special status of time remains,  so that the anisotropic scaling  between space and time,
\bq
\lb{1.1}
{\bf x} \rightarrow b^{-1} {\bf x}, \;\;\;  t \rightarrow b^{-3} t,
\eq
can  still be  realized,  and the theory is UV complete. Meanwhile, because of the elimination of the spin-0 graviton,  its IR  behavior can be
 significantly improved.
The elimination of the spin-0 graviton was done  initially in the   case $\lambda = 1$ \cite{HMT,WW}, but soon generalized
to the case with any $\lambda$ \cite{Silva,HW,LWWZ}, where $\lambda$ denotes a coupling constant that
characterizes the deviation of  the kinetic part of the action from the corresponding one given in GR.
 (For the analysis of the Hamiltonian structure of the theory, see \cite{HMT,Kluson}).
 
 Under the coordinate transformations (\ref{1.4}), the lapse function $N$, the shift vector $N_{i}$, the 3-metric $g_{ij}$,
 the $U(1)$ gauge field $A$ and the Newtonian prepotentail $\varphi$ transform, reslectively, as
 \bqn
 \lb{1.2}
 \delta{N} &=& \zeta^{k}\nabla_{k}N + \dot{N}f + N\dot{f},\nb\\
\delta{N}_{i} &=& N_{k}\nabla_{i}\zeta^{k} + \zeta^{k}\nabla_{k}N_{i}  + g_{ik}\dot{\zeta}^{k}
+ \dot{N}_{i}f + N_{i}\dot{f}, \nb\\
\delta{g}_{ij} &=& \nabla_{i}\zeta_{j} + \nabla_{j}\zeta_{i} + f\dot{g}_{ij}, \nb\\
\delta{A} &=& \zeta^{i}\partial_{i}A + \dot{f}A  + f\dot{A},\nb\\
\delta \varphi &=&  f \dot{\varphi} + \zeta^{i}\partial_{i}\varphi,
\eqn
where $\dot{f} \equiv df/dt,\;  \nabla_{i}$ denotes the covariant 
derivative with respect to   $g_{ij}$,  and $\delta{g}_{ij} 
\equiv \tilde{g}_{ij}\left(t, x^k\right) - {g}_{ij}\left(t, x^k\right)$,
 etc. From these expressions one can see that   $N$ and   $N_{i}$ play the role of gauge fields of the Diff($M, \; {\cal{F}}$). 
 Therefore, it is natural to assume that $N$ and $N_{i}$ inherit the same dependence on 
space and time as the corresponding generators \cite{Horava}, 
\bq
\lb{1.6}
N = N(t), \;\;\; N_{i} = N_{i}(t, x),
\eq
which is  often referred to as the projectability condition. On the other hand, under the U(1) gauge transformation, the above quantities transform as
\bqn
\lb{1.3}
\delta_{\alpha}{N} &=& 0,\;\;\;
\delta_{\alpha}N_{i}  = N\nabla_{i}\alpha,\;\;\;
\delta_{\alpha}g_{ij} = 0,\nb\\
\delta_{\alpha}A &=&\dot{\alpha} - N^{i}\nabla_{i}\alpha,\;\;\;
\delta_{\alpha}\varphi = - \alpha,
\eqn
where $\alpha [= \alpha(t, x)]$ is   the generator  of the local $U(1)$ gauge symmetry, and  $N^{i} = g^{ik}N_{k}$. For the detail, we refer readers to \cite{HMT,WW}. 

 It should be  noted  that all the above hold only in the case with projectability condition \cite{Horava,reviews}.
However,   the elimination of the spin-0 graviton can be also realized in the non-projectability
 case with the extended  symmetry (\ref{symmetry}) \cite{ZWWS,ZSWW}. In addition, the number of independent coupling constants
 can be significantly reduced (from more than 70 \cite{BPS,KP} to 15), by simply imposing  the  softly breaking detailed balance condition,  while
the theory  still remains UV complete and has  a healthy IR limit.  When applying it to cosmology, a remarkable result is obtained:
 the Friedmann-Robertson-Walker universe is necessarily flat.  %For detail, we refer readers to \cite{ZWWS}.

 In this paper, we study electromagnetic  static spacetimes   in the Ho\v{r}ava-Melby-Thompson (HMT) setup  \cite{HMT}, in which
 $\lambda = 1$ and the projectability condition (\ref{1.6}) is adopted.
 Specifically,   after giving a  brief introduction to the HMT theory in Sec. II,  we consider its coupling to a vector field in Sec. III, while
 in Sec. IV we study electric static spacetimes, and find all the static solutions. In Sec. V we study the existence of horizons and show that
 some of these solutions found in Sec. IV have black hole structures  in the geometrical optical approximations  \cite{GLLSW}. It should be noted that
 because of the breaking of the general covariance, $\tilde{x}^{\mu} = \zeta^{\mu}({t}, {x}), (\mu = 0, 1, 2, 3)$,    the dispersion relations of test  particles,
 including the massless ones,  contain generically high-order momentum terms, so the speeds of these particles are unbounded 
in the ultraviolet (UV), and the causal structures of spacetimes are quite different from that presented in  GR. In particular, 
in the HL theory the conception  of light-cones is not valid  any longer,  and the motions of test particles do not follow geodesics  \cite{GLLSW}. As a result, the definitions of black holes 
given in GR \cite{HE73,Tip77,Hay94,Wang} cannot be applied to the HL gravity without modifications.  In Sec. VI  we present our main conclusions.

It should be noted that electromagnetic  static spacetimes in other versions of the HL theory were studied in \cite{EM}, while a new mechanism
for generation of primordial magnetic seed field in the early universe   without the local U(1) symmetry  was   considered in \cite{MMS}.

\section{Nonrelativisitc general covariant theory }

\renewcommand{\theequation}{2.\arabic{equation}} \setcounter{equation}{0}

In this section, we   give a very brief introduction to the nonrelativistic general covariant theory of gravity, proposed recently by
HMT. For details, we refer readers to  \cite{HMT,WW}. We shall closely follow \cite{WW}, so that the notations and conversations will be
used directly from there without further explanations. %In the following, \cite{WW} will be referred to as Paper I.

The basic variables are $A, \; \varphi, \; N, \; N_{i}$ and $g_{ij}$, %where $N$ is the lapse function, $N_{i}$ the shift vector, and $g_{ij}$ the 3-metric,
in terms of which the spacetime  is given by,
\bq
\lb{2.0}
ds^2= -N^2c^{2}dt^2+g_{ij}(dx^i+N^idt)(dx^j+N^jdt).
\eq
The total action takes the form,
 \bqn \lb{2.4}
S &=& \zeta^2\int dt d^{3}x N \sqrt{g} \Big({\cal{L}}_{K} -
{\cal{L}}_{{V}} +  {\cal{L}}_{{\varphi}} +  {\cal{L}}_{{A}} \nb\\
& & ~~~~~~~~~~~~~~~~~~~~~~ \left. +\frac{1}{\zeta^{2}} {\cal{L}}_{M} \right),
 \eqn
where $g={\rm det}\,g_{ij}$, and
 \bqn \lb{2.5}
{\cal{L}}_{K} &=& K_{ij}K^{ij} -   K^{2},\nb\\
{\cal{L}}_{\varphi} &=&\varphi {\cal{G}}^{ij} \Big(2K_{ij} + \nabla_{i}\nabla_{j}\varphi\Big),\nb\\
{\cal{L}}_{A} &=&\frac{A}{N}\Big(2\Lambda_{g} - R\Big).
 \eqn
Here   the coupling constant $\Lambda_{g}$, acting like a 3-dimensional cosmological
constant, has the dimension of (length)$^{-2}$. %$\nabla_{i}$ denotes the covariant derivatives with respect to the 3-metric $g_{ij}$.
 $K_{ij}$ is the extrinsic curvature of the hypersurfaces $t = $ Constant, and ${\cal{G}}_{ij}$ is the 3-dimensional ``generalized"
Einstein tensor, defined, respectively,  by
 \bqn \lb{2.6}
K_{ij} &=& \frac{1}{2N}\left(- \dot{g}_{ij} + \nabla_{i}N_{j} +
\nabla_{j}N_{i}\right),\nb\\
{\cal{G}}_{ij} &=& R_{ij} - \frac{1}{2}g_{ij}R + \Lambda_{g} g_{ij},
 \eqn
 where  the Ricci  tensor $R_{ij}$  refers to the three-metric $g_{ij}$, and $R [= g^{ij}R_{ij}]$ denotes the 3D Ricci scalar.
% where %$N_{i} = g_{ij}N_{j}$.
${\cal{L}}_{M}$ is the
matter Lagrangian  and
${\cal{L}}_{{V}}$  an arbitrary Diff($\Sigma$)-invariant local scalar functional
built out of the spatial metric $g_{ij}$, its Riemann tensor and spatial covariant derivatives, without the use of time derivatives.
In \cite{SVW}, by assuming that the highest order derivatives are six and that  the theory  respects
the parity and time-reflection symmetry,  the most general form of  ${\cal{L}}_{{V}}$ is given by \cite{SVW} (See also \cite{KK}),
 \bqn \lb{2.5a}
{\cal{L}}_{{V}} &=& \zeta^{2}g_{0}  + g_{1} R + \frac{1}{\zeta^{2}}
\left(g_{2}R^{2} +  g_{3}  R_{ij}R^{ij}\right)\nb\\
& & + \frac{1}{\zeta^{4}} \left(g_{4}R^{3} +  g_{5}  R\;
R_{ij}R^{ij}
+   g_{6}  R^{i}_{j} R^{j}_{k} R^{k}_{i} \right)\nb\\
& & - \frac{1}{\zeta^{4}} \left[g_{7}R\nabla^{2}R +  g_{8}
\left(\nabla_{i}R_{jk}\right)
\left(\nabla^{i}R^{jk}\right)\right],  ~~~~
 \eqn
 where the coupling  constants $ g_{s}\, (s=0, 1, 2,\dots 8)$  are all dimensionless, and
 \bq
 \lb{lamda}
 \Lambda = \frac{1}{2}\zeta^{2}g_{0},
 \eq
 denotes the cosmological constant. The relativistic limit in the IR
 requires,
 \bq
 \lb{2.5b}
 g_{1} = -1,\;\;\; \zeta^2 = \frac{1}{16\pi G},
 \eq
where $G$ denotes the Newtonian constant.

Variation of the total action (\ref{2.4}) with respect to the lapse function $N(t)$  yields the
Hamiltonian constraint,
 \bq \lb{eq1}
\int{ d^{3}x\sqrt{g}\left({\cal{L}}_{K} + {\cal{L}}_{{V}} - \varphi {\cal{G}}^{ij}\nabla_{i}\nabla_{j}\varphi\right)}
= 8\pi G \int d^{3}x {\sqrt{g}\, J^{t}},
 \eq
where
 \bq \lb{eq1a}
J^{t} = 2 \frac{\delta\left(N{\cal{L}}_{M}\right)}{\delta N}.
 \eq

Variation of the action with respect to the shift vector $N_{i}$ yields the
supermomentum constraint,
 \bq \lb{eq2}
\nabla_{j}\Big(\pi^{ij} - \varphi  {\cal{G}}^{ij}\Big) = 8\pi G J^{i},
 \eq
where the supermomentum $\pi^{ij} $ and matter current $J^{i}$
are defined as
 \bqn \lb{eq2a}
\pi^{ij} &\equiv& %\frac{\delta{\cal{L}}_{K}}{\delta\dot{g}_{ij}}
  - K^{ij} +  K g^{ij},\nb\\
J^{i} &\equiv& - N\frac{\delta{\cal{L}}_{M}}{\delta N_{i}}.
 \eqn
Similarly, variations of the action with respect to $\varphi$ and $A$ yield, respectively,
\bqn
\lb{eq4a}
& & {\cal{G}}^{ij} \Big(K_{ij} + \nabla_{i}\nabla_{j}\varphi\Big) = 8\pi G J_{\varphi},\\
\lb{eq4b}
& & R - 2\Lambda_{g} =    8\pi G J_{A},
\eqn
where
\bq
\lb{eq5}
J_{\varphi} \equiv - \frac{\delta{\cal{L}}_{M}}{\delta\varphi},\;\;\;
J_{A} \equiv 2 \frac{\delta\left(N{\cal{L}}_{M}\right)}{\delta{A}}.
\eq
On the other hand, variation with respect to $g_{ij}$ leads to the
dynamical equations,
 \bqn
  \lb{eq3}
&&
\frac{1}{N\sqrt{g}}\left[\sqrt{g}\left(\pi^{ij} - \varphi {\cal{G}}^{ij}\right)\right]_{,t} %^{\displaystyle{\cdot}}
= -2\left(K^{2}\right)^{ij}+2K K^{ij}
\nb\\
& &  ~~~~~ + \frac{1}{N}\nabla_{k}\left[N^k \pi^{ij}-2\pi^{k(i}N^{j)}\right]\nb\\
& & ~~~~~
+  \frac{1}{2} \left({\cal{L}}_{K} + {\cal{L}}_{\varphi} + {\cal{L}}_{A}\right) g^{ij} \nb\\
& &  ~~~~~    + F^{ij} + F_{\varphi}^{ij} +  F_{A}^{ij} + 8\pi G \tau^{ij},
 \eqn
where  $f_{(ij)} \equiv \left(f_{ij} + f_{ji}\right)/2$, and
 \bqn
\lb{eq3a}
 \left(K^{2}\right)^{ij} &\equiv& K^{il}K_{l}^{j}, \nb\\
F_{A}^{ij} &\equiv& \frac{1}{N}\left[AR^{ij} - \Big(\nabla^{i}\nabla^{j} - g^{ij}\nabla^{2}\Big)A\right],\nb\\
F^{ij} &\equiv& - \frac{1}{\sqrt{g}}\frac{\delta\left(\sqrt{g} {\cal{L}}_{V}\right)}{\delta{g}_{ij}}
 = \sum^{8}_{s=0}{g_{s} \zeta^{n_{s}} \left(F_{s}\right)^{ij} },\nb\\
 F_{\varphi}^{ij} &\equiv&  \sum^{3}_{n=1}{F_{(\varphi, n)}^{ij}},
 \eqn
with %The constants are given by $g_{0} = {2\Lambda}{\zeta^{-2}}$, $g_{1} = -1$, and
$n_{s} =(2, 0, -2, -2, -4, -4, -4, -4,-4)$.  The 3-tensors $ \left(F_{s}\right)_{ij}$ and
$F_{(\varphi, n)}^{ij}$ are given in Appendix A. The stress 3-tensor $\tau^{ij}$ is defined as
 \bq \label{tau}
\tau^{ij} = {2\over \sqrt{g}}{\delta \left(\sqrt{g}
 {\cal{L}}_{M}\right)\over \delta{g}_{ij}}.
 \eq

The matter, on the other hand, satisfies  the conservation laws,
 \bqn \lb{eq5a} & &
 \int d^{3}x \sqrt{g} { \left[ \dot{g}_{kl}\tau^{kl} -
 \frac{1}{\sqrt{g}}\left(\sqrt{g}J^{t}\right)_{, t}
 +   \frac{2N_{k}}  {N\sqrt{g}}\left(\sqrt{g}J^{k}\right)_{,t}
  \right.  }   \nb\\
 & &  ~~~~~~~~~~~~~~ \left.   - 2\dot{\varphi}J_{\varphi} -  \frac{A} {N\sqrt{g}}\left(\sqrt{g}J_{A}\right)_{,t}
 \right] = 0,\\
\lb{eq5b} & & \nabla^{k}\tau_{ik} -
\frac{1}{N\sqrt{g}}\left(\sqrt{g}J_{i}\right)_{,t}  - \frac{J^{k}}{N}\left(\nabla_{k}N_{i}
- \nabla_{i}N_{k}\right)   \nb\\
& & \;\;\;\;\;\;\;\;\;\;\;- \frac{N_{i}}{N}\nabla_{k}J^{k} + J_{\varphi} \nabla_{i}\varphi - \frac{J_{A}}{2N} \nabla_{i}A
 = 0.
\eqn

\section{Coupling of a Vector Field}

\renewcommand{\theequation}{3.\arabic{equation}} \setcounter{equation}{0}

To couple the gravitational sector ($N, N_{i}, g_{ij}, A, \varphi$) with a vector field, we borrow the recipe of  \cite{Silva},
in which it was shown that for any given matter field, say, $\psi_{n}$, that is invariant under ${\mbox{Diff}}(M, {\cal{F}})$, its motion is described by the
action, $\hat{S}_{M}(N, N_{i}, g_{jk}; \psi_{n})$. Then, the action,
\bqn
\lb{3.1}
&& S_{M}\left(N, N_{i}, g_{jk}, A, \varphi; \psi_{n}\right)  = \hat{S}_{M}\left(N, \hat{N}_{i}, g_{jk}; \psi_{n}\right)\nb\\
& &
~~~~~~~~~~ + \int{dtd^3x N\sqrt{g} Z(g_{ij}, \psi_{n})\big(A - {\cal{A}})},
\eqn
has the enlarged symmetry (\ref{symmetry}), where $Z(g_{ij}, \psi_{n})$ denotes the most general
scalar operator of dimension two, $[Z] = 2$,  of  the enlarged symmetry, and
\bqn
\lb{3.2}
\hat{N}_{i} &\equiv& N_{i} + N\nabla_{i}\varphi,\nb\\
 {\cal{A}} &\equiv&  - \dot{\varphi} + N^{k}\nabla_{k}\varphi + \frac{1}{2}N \big(\nabla_{k}\varphi\big)  \big(\nabla^{k}\varphi\big).
 \eqn

In \cite{KK},   the general action of a massive vector field  $(A_{0}, A_{i})$ with the  ${\mbox{Diff}}(M, {\cal{F}})$ symmetry (\ref{1.4})  was constructed. Applying the above recipe
to this vector field, we obtain   the action of a massive vector field that is invariant under  the enlarged symmetry (\ref{symmetry}),
\bqn
\lb{3.3}
{S}_{EM}&=& \frac{1}{4g_e^2}\int{dt d^{3}x \sqrt{g}\Bigg[\frac{2}{N}g^{ij}({\cal{F}}_{0i}-\hat{N}^k{\cal{F}}_{ki})}\nb\\
&& \times ({\cal{F}}_{0j}-\hat{N}^l{\cal{F}}_{jl}) {+\frac{m_{e}^2}{N}(A_{0}-\hat{N}^iA_i)^2}\nb\\
&&-N{\cal{G}}   {+4g_{e}^2{\cal{K}} B_iB^i(A-\cal{A})\Bigg] },
\eqn
where $m_{e}$ denotes the mass of the vector field, ${\cal{K}}$ is an arbitrary function of $A^{k}A_{k}$,
%${\cal{F}}_{0i} = \partial_{i}A_{0} - \partial_{t}A_{i} $,
%${\cal{F}}_{ij} = \partial_{j}A_{i} - \partial_{i}A_{j}$,
and
\bqn
\lb{3.4}
{\cal{F}}_{0i} &=& \partial_{i}A_{0} - \partial_{t}A_{i},\;\;\;{\cal{F}}_{ij} = \partial_{j}A_{i} - \partial_{i}A_{j},\nb\\
  B_i &=& \frac{1}{2}\frac{\varepsilon_i^{\;\;jk}}{\sqrt{g}}{\cal{F}}_{jk}, \;\; %F_{ij}=\frac{\varepsilon_{ij}^k}{\sqrt{g}}B_k,
  \nabla^iB_i=0,\nb\\
{\cal{G}}  &=& a_0+a_1\zeta_1+a_2\zeta_1^2+a_3\zeta_1^3+a_4\zeta_2+a_5\zeta_1\zeta_2\nb\\
&& +a_6\zeta_3+a_7\zeta_4,
\eqn
with $a_{n}$ being arbitrary functions of  $A^{k}A_{k}$ only, and
\bqn
\lb{3.5}
  \zeta_1&=& B_iB^i, \;\;\;  \zeta_2=\left(\nabla_i B_j\right)\left(\nabla^i B^j\right), \nb\\
  \zeta_3&=& \left(\nabla_i B_j\right)\left(\nabla^i B^k\right)\left(\nabla^j B_k\right),\nb\\
    \zeta_4&=& \left(\nabla_i \nabla_j B_k\right)\left(\nabla^i \nabla^j B^k\right).
\eqn
In terms of the magnetic field $B_{i}$, the 3-tensor ${\cal{F}}_{ij}$ can be written as
\bq
\lb{3.6}
{\cal{F}}_{ij}\equiv \frac{\varepsilon^k_{\;\;ij}}{\sqrt{g}}B_k.
\eq

 Variations of $S_{EM}$ with respect to $A_{0}$ and $A_i$ yield the generalized Maxwell equations, given respectively, by
\bqn
\lb{3.7a}
&& \nabla^k{\cal{F}}_{0k} = \frac{1}{2}  m_{e}^2\left(A_0-\hat{N}^iA_i\right),\\
\lb{3.7b}
&& \frac{1}{N\sqrt{g}}\partial_t \left[\frac{\sqrt{g}}{N}g^{ij}\left({\cal{F}}_{0j}-\hat{N}^l{\cal{F}}_{jl}\right)\right]=- \frac{1}{N^2}\Bigg\{g^{kj}\nabla_k\Big[\hat{N}^i \nb\\
&&  \times\big({\cal{F}}_{0j}-\hat{N}^l{\cal{F}}_{jl}\big)\Big]-g^{ij}\nabla_k\Big[\hat{N}^k\big({\cal{F}}_{0j}-\hat{N}^l{\cal{F}}_{jl}\big)\Big]\Bigg\} \nb\\
&& +\frac{m_{e}^2}{2N^2}\big(A_0 -\hat{N}^jA_j\big)\hat{N}^i+\frac{1}{2}A^i\Big(a'_0 + a'_1 \zeta_1+ a'_2 \zeta_1^2\nb\\
&& + a'_3 \zeta_1^3+ a'_4 \zeta_2+ a'_5 \zeta_1\zeta_2+ a'_6 \zeta_3+ a'_7 \zeta_4 \Big)\nb\\
&& +\frac{\varepsilon_k^{\;\;ji}}{2\sqrt{g}}\nabla_j\Big[\big(a_1+2a_2\zeta_1+3a_3\zeta_1^2+a_5\zeta_2\big)B^k\Big]\nb\\
&& +\frac{\varepsilon_j^{\;\;ik}}{4\sqrt{g}}\nabla_k \nabla_l\Bigg\{2\left(a_4 +a_5\zeta_1\right) \nabla^lB^j\nb\\
&& +a_6 \Big[\left(\nabla^lB^m\right) \left(\nabla^jB_m\right) + \left(\nabla^l B_m\right)\left(\nabla^mB^j\right)\nb\\
&& +\left(\nabla_mB^l\right)\left(\nabla^mB^j\right)\Big]\Bigg\}\nb\\
&& +\frac{\varepsilon_k^{\;\;li}}{2\sqrt{g}}\nabla_l \nabla_j\nabla_m\Big(a_7\nabla^m\nabla^jB^k\Big)\nb\\
&& - \frac{2g_e^2}{N}{\cal{K}}'A^iB_jB^j(A-\cal{A})\nb\\
&& +\frac{g_e^2\varepsilon_j^{\;\;ik}}{N\sqrt{g}}\nabla_k\Big[{\cal{K}}B^j(A-\cal{A})\Big]\nb\\
&& -\frac{g_e^2\varepsilon_k^{\;\;ji}}{N\sqrt{g}}\nabla_j\Big[{\cal{K}}B^k(A-\cal{A})\Big],
\eqn
where a prime denotes the ordinary derivative with respect to the indicated argument.

On the other hand, when the electromagnetic field is the only source, we find that $J^{t}, J_{i}, J_{\varphi}, J_{A}$ and $\tau_{ij}$ are given by 
Eq.(\ref{3.8}) in Appendix B. 

\section{ Spherical  static spacetimes filled with an electromagnetic field}

\renewcommand{\theequation}{4.\arabic{equation}} \setcounter{equation}{0}

Spherically symmetric static vacuum spacetimes with projectability condition in the HMT setup were  studied systematically in \cite{GPW,GSW,GLLSW,AP}.
In particular, the metric %for  static spherically symmetric spacetimes that preserve the ADM form %of Eq. (\ref{1.2})with the projectability condition
can be cast in the form,
\bq
\lb{4.1}
ds^{2} = - c^{2}dt^{2} + e^{2\nu} \left(dr + e^{\mu - \nu} dt\right)^{2}  + r^{2}d^2\Omega,
\eq
in the spherical coordinates $x^{i} = (r, \theta, \phi)$, where  $d^2\Omega = d\theta^{2}  + \sin^{2}\theta d\phi^{2}$,
and
\bq
\lb{4.2}
 \mu = \mu(r),\;\;\; \nu = \nu(r),\;\;\; N^{i} = e^{\mu - \nu}\delta^{i}_{r}.
 \eq
The corresponding timelike Killing vector is  $\xi = \partial_{t}$.
%we leave the choice of the $U(1)$ gauge open. From
% Eq.(\ref{2.3}) one can see that it can be used to set  one (and only one) of the three functions $A,\; \varphi$
% and $N_{r}$  to zero.  To compare our results  with the one obtained in \cite{HMT}, in the
% rest of this paper (except the first part of Sec. VII),
With the gauge freedom of the Newtonian prepotential and the electromagnetic field, without loss of the generality, we  choose the gauges,
 \bq
 \lb{gauge}
 \varphi =0,\;\; A_{i}(r) = A_{1}(r)\delta_{i}^{r},
 \eq
that is,   we consider only the electric field and set the magnetic field to zero.   Then,   we find that
\bqn
\lb{4.3}
{\cal{L}}_{\varphi} &=& 0,\;\;\; F_{\varphi}^{ij} = 0,\;\;\; B_{i} = 0,\nb\\
\pi_{ij} &=&-e^{\mu+\nu}\Big(\mu'\delta_i^r\delta_j^r+re^{-2\nu}\Omega_{ij}\Big)\nb\\
& & +e^{\mu-\nu}g_{ij}\Big(\mu'+\frac{2}{r}\Big),\nb\\
K_{ij} &=& e^{\mu+\nu}\Big(\mu'\delta^{r}_{i}\delta^{r}_{j} + re^{-2\nu}\Omega_{ij}\Big),\nb\\
R_{ij} &=&  \frac{2\nu'}{r}\delta^{r}_{i}\delta^{r}_{j} + e^{-2\nu}\Big[r\nu' - \big(1-e^{2\nu}\big)\Big]\Omega_{ij},\nb\\
{\cal{L}}_{K} &=& - \frac{2}{r^{2}} e^{2(\mu-\nu)}\left(2r\mu' + 1\right),  \nb\\
{\cal{L}}_{A} &=&  \frac{2A}{r^2} \Big[e^{-2\nu}\left(1 - 2r \nu'\right) + \left(\Lambda_{g} r^2 - 1\right)\Big],
%{\cal{L}}_{V} &=& \sum_{s=0}^{3}{{\cal{L}}_{V}^{(s)}},
\eqn
where   $\Omega_{ij} \equiv \delta^{\theta}_{i}\delta^{\theta}_{j}  + \sin^{2}\theta\delta^{\phi}_{i}\delta^{\phi}_{j}$,
  and ${\cal{L}}_{V}$ is too complicated to be given explicitly here. %s are given by Eq.(A1) in \cite{GPW}.
 Then,  the Hamiltonian constraint (\ref{eq1}) reads,
 \bq
 \lb{4.4}
\int{\left( {\cal{L}}_{K} + {\cal{L}}_{V}- 8 \pi G J^{t} \right) e^{\nu} r^{2} dr}
= 0,
 \eq
 where
 \bq
\lb{4.5}
J^t =  - \frac{1}{2g_e^2}\Big[2A_0'^{2}e^{-2\nu} +m_{e}^{2}\left(A_{0} - A_{1}  e^{\mu-\nu}\right)^2\Big].
\eq
The momentum constraint (\ref{eq2}) reduces to,
 \bqn
 \lb{4.6}
 \nu' e^{\mu} &=& - \frac{2\pi Gm_{e}^2}{g_e^2}\Big(A_0e^{\nu}-A_1e^{\mu}\Big)rA_1,
 \eqn
 where
 \bq
 \lb{4.7}
 J^{i} =    \frac{m_{e}^2}{2g_e^2}(A_0-A_1e^{\mu-\nu})A_1e^{-2\nu} \delta^{i}_{r}.
 \eq
 Eqs.(\ref{eq4a}) and (\ref{eq4b}), on the other hand,  now read, respectively,
 \bqn
 \lb{4.8a}
& &  \Big[e^{2\nu}\left(\Lambda_{g}r^2 -1\right) + 1\Big]e^{\mu + \nu}\mu'-2\Big(\nu' - \Lambda_{g}re^{2\nu}\Big)e^{\mu + \nu}\nb\\
&&= 8\pi G r^{2} e^{4\nu} J_{\varphi}, ~~~~~~~~ \\
\lb{4.8b}
& & 2 r \nu'  - \Big[e^{2\nu}\left(\Lambda_{g}r^2 - 1\right) +  1\Big] =0,
\eqn
where $J_A = 0$, and
\bqn
\lb{4.8c}
J_\varphi &=&\frac{m_{e}^2e^{-2\nu}}{2g_e^2}\Bigg[(A_1A_0)' - 2 A_1 A'_1e^{\mu-\nu}\nb\\
&&- A_1^2e^{\mu-\nu}\left(\mu'-2\nu'+\frac{2}{r}\right)\nb\\
&& - A_0A_1\left(\nu'-\frac{2}{r}\right)\Bigg].
\eqn
 The dynamical equations (\ref{eq3}) yield,
\bqn
 \lb{4.9a}
& &e^{2\mu}\left[2\big(\mu' + \nu'\big) + \frac{1}{r}\right] + \frac{1}{2}re^{2\nu}{\cal{L}}_{A}= - r\Big(F_{rr}  \nb\\
 & &   + F^{A}_{rr} + 8\pi G \tau_{rr} \Big),~~~~\\
 \lb{4.9b}
 & &  e^{2\mu}\left[\mu'' + \big(2\mu' - \nu'\big)\left(\mu' + \frac{1}{r}\right)\right]  + \frac{1}{2}e^{2\nu}{\cal{L}}_{A} \nb\\
 & &  ~  = -\frac{e^{ 2\nu }}{r^{2}}\Bigg(F_{\theta\theta}  + F^{A}_{\theta\theta} + 8\pi G \tau_{\theta \theta} \Bigg), ~~~~
 \eqn
 where
 \bqn
\lb{4.10}
 \tau_{rr}   &=&- \frac{1}{4g_e^2}\Big[{\cal{G}} e^{2\nu}+2A_0'^{2}-2 a'_0A_1^2 \nb\\
 &&~~~~~~~~~ - m^{2}_{e}e^{2\nu}\left(A_{0} - A_{1}e^{\mu-\nu}\right)^{2}\Big],\nb\\
\tau_{\theta \theta} &=& - \frac{r^{2}e^{-2\nu}}{4g_e^2}\Big[{\cal{G}} e^{2\nu}- 2A_0'^{2}
\nb\\ &&~~~~~~~~~~~~~
- m^{2}_{e}e^{2\nu}\left(A_{0} - A_{1}e^{\mu-\nu}\right)^{2}\Big].~~~~
 \eqn
The Maxwell equations (\ref{3.7a}) and (\ref{3.7b})  now become,
\bqn
\lb{4.11a}
&& a'_{0}A_{1} + m_{e}^{2}e^{\mu+\nu}\left(A_0 - A_1e^{\mu-\nu}\right) = 0,\\
\lb{4.11b}
& & A_0'' - A_0'\nu'+ \frac{2}{r}A_0' - \frac{1}{2}m_{e}^2e^{2\nu}\big(A_0-e^{\mu-\nu}A_1\big)=0. ~~~~~~~~~~
 \eqn

\subsection{Massless Electromagnetic Field with $N^{r} \not=0$}

When $N^{r} \not= 0$ (or $\mu \not= -\infty$), for a massless electromagnetic field, Eq.(\ref{4.6}) immediately gives
$\nu = \nu_{0}$, where $\nu_{0}$ is an integration constant. Then, Eq.(\ref{4.8b}) yields,
\bq
\lb{4.12}
\nu = 0 = \Lambda_{g}.
\eq
When $m_{e} = 0$, Eqs.(\ref{4.11a}) and (\ref{4.11b}) yield,
\bq
\lb{4.13}
A_{0} = \frac{Q}{r} + Q_{0},\;\;\; A_{1} = 0,
\eq
where $Q$ and $Q_{0}$ are two integration constant. Without loss of generality, we can always set $Q_{0} = 0$. It is remarkable that
the above solution for the electromagnetic field is the same as that given in GR.
On the other hand, when $\nu = 0$, the spatial part is flat, $R_{ij} = 0$, and all the high order spatial derivative terms vanish, so we have
$F^{ij} = -\Lambda g_{ij}$. Then, Eqs.(\ref{4.9a}) and (\ref{4.9b}) reduce to,
\bqn
\lb{4.14a}
&& \left(2\mu' + \frac{1}{r}\right)e^{2\mu}  =\Lambda r^{2} - 2r A' + \frac{4\pi G Q^{2}}{g_{e}^{2}r^{2}},\\
\lb{4.14b}
&& \left[\mu'' + 2\mu'\left(\mu' + \frac{1}{r}\right)\right]e^{2\mu} =  \frac{1}{r}\Bigg[\Lambda r - \left(rA'\right)'   \nb\\
&& ~~~~~~~~~~~~~~~~~~~~~~~~~~~~ - \frac{4\pi G Q^{2}}{g_{e}^{2}r^{3}}\Bigg].
\eqn
Note that these two equations are not independent. In fact, one can obtain Eq.(\ref{4.14b}) from Eq.(\ref{4.14a}).
Therefore, we have one equation for two unknowns, $\mu$ and $A$.
Thus, similar to the vacuum case \cite{GSW}, for any chosen gauge field $A$, the metric coefficient $\mu$ is given by
\bqn
\lb{4.15}
\mu &=& \frac{1}{2}\ln\Bigg(\frac{2m}{r} + \frac{1}{3}\Lambda r^{2}  - \frac{4\pi G Q^{2}}{g_{e}^{2}r^{2}}\nb\\
&& ~~~~~~~~~~ - 2A + \frac{2}{r}\int^{r}{A(r')dr'}\Bigg). %\right\}.
\eqn

Then,  the Hamiltonian constraint (\ref{4.4}) reduces to
\bq
\lb{4.16}
\int_0^\infty{A'r dr} = 0.
\eq
Therefore, for any given $A$,  the solutions of Eqs.(\ref{4.12}), (\ref{4.13}) and (\ref{4.15}) represent solutions of the HL theory in the HMT setup,  coupled with
an electromagnetic field,   provided that Eq.(\ref{4.16}) is satisfied.

When $A = $ Constant, the above solutions  reduce exactly to the Reissner-Nordstr\"om solution found in GR but written in  
 the Painleve-Gullstrand coordinates \cite{GP}. %Arnowitt-Deser-Misner form (\ref{4.1}).

\subsection{Massless Electromagnetic Field with $N^{r} =0$}

In the diagonal case, we have
\bq
\lb{NR}
 N^{r} = 0, \;\;\; {\mbox{or}} \;\;\; \mu = -\infty.
 \eq
Then,  $K_{ij}=0 = \pi_{ij}$, for which  the momentum constraint is satisfied identically, while  Eq.(\ref{4.8b}) becomes,
\bq
\lb{4.71a}
\nu'=\frac{1}{2r}\Big[e^{2\nu}(\Lambda_{g}r^2-1)+1\Big],
\eq
which has the general solution,
\bq
\lb{4.71b}
 \nu = - \frac{1}{2}\ln\left(1 - \frac{2M}{r} -\frac{\Lambda_{g}}{3}r^2\right),
\eq
where $M$ is a constant. 
Then, the Maxwell equations (\ref{4.11a}) and (\ref{4.11b}) reduce to,
\bqn
\lb{4.71c}
&& A_1 = 0,\\
\lb{4.71ca}
&& A_0''+A_0'\Big(\frac{2}{r}-\nu'\Big) = 0.
\eqn
Eq.(\ref{4.71ca}) has the general solution, %which yield, %Integrating of the latter the first equation turns $A_0$ to be
\bqn
\lb{4.71d}
&& A_0 = D_1\int\frac{dr}{r^{4}\left(1  - \frac{2M}{r} -\frac{\Lambda_{g}}{3}r^2\right)}+D_2,
\eqn
where $D_1$ and $D_2$ are two integration constants. For the solutions   $\mu$ and $\nu$, given by Eqs.(\ref{NR}) and (\ref{4.71b}), it can be shown that
only one of the  two dynamical equations  (\ref{4.9a}) and (\ref{4.9b}) is independent, and can be cast in the form,

\bq
\lb{4.71e}
A' + P(r)A = Q(r),
%&& A'-\frac{B}{r(r-2B)}A = \frac{g_0\zeta^2r^2}{4(r-2B)}\nb\\
%&& -\frac{B}{r(r-2B)}+\frac{g_3\zeta^{-2}B^2}{2(r-2B)r^4}\nb\\
%&& -\frac{6B^2\zeta^{-4}g_5}{r-2B}\Bigg(\frac{-6}{r^6}+\frac{11B}{r^7}\Bigg)\nb\\
%&& +\frac{3\zeta^{-4}g_6B^2}{2(r-2B)}\Bigg(\frac{27}{r^6}-\frac{50B}{r^7}\Bigg)\nb\\
%&& +\frac{3\zeta^{-4}g_8B^2}{2(r-2B)}\Bigg(\frac{-21}{r^6}+\frac{40B}{r^7}\Bigg)\nb\\
%&& -\frac{D_1^2}{2g_e^2r^2}.
\eq
where
\bqn
\lb{4.71ea}
P(r) &=& \frac{ \Lambda_{g}r^{3}  -3M}{r\left[3(r- 2M) - \Lambda_{g}r^{3}\right]},\nb\\
Q(r) &=& \frac{1}{2\left[3(r- 2M) - \Lambda_{g}r^{3}\right]}\left(\frac{4\pi G D^{2}_{1}}{g_{e}^{2} r^{2}}  \right.\nb\\
&& \left.  - \alpha_{1}r^{2}
- \frac{\alpha_{2}}{r}  - \frac{\alpha_{3}}{r^{4}} - \frac{\alpha_{4}}{r^{6}} - \frac{\alpha_{5}}{r^{7}}\right),
\eqn
with
\bqn
\lb{4.71eb}
 \alpha_1 &=& -\frac{3}{2}g_0\zeta^2+\Lambda_{g}+\left(2g_2+\frac{2}{3}g_3\right)\Lambda_{g}^2\zeta^{-2} \nb\\
                 && + \left(12g_4+4g_5+\frac{4}{3}g_6\right)\Lambda_{g}^3\zeta^{-4}, \nb\\
 \alpha_2  &=&  6M - \big(24g_2+10g_3\big)M\Lambda_{g}\zeta^{-2}\nb\\
                   && - \big(72g_4+28g_5+12g_6-2g_8\big)M\Lambda_{g}^2\zeta^{-4},\nb\\
 \alpha_3  &=&  -3g_3M^2 \zeta^{-2}+ \big(78g_5+90g_6-75g_8\big)M^2\Lambda_{g}\zeta^{-4},\nb\\
 \alpha_4  &=&  27M^2\zeta^{-4}\big(7g_8-9g_6-8g_5\big),\nb\\
 \alpha_5  &=&  -18M^3\zeta^{-4}\big(20g_8-25g_6-22g_5\big).
\eqn
The general solution of Eq.(\ref{4.71e})  is given by,
\bqn
\lb{4.71ec}
A(r) &=& e^{-\int^{r}{P(r')dr'}}\nb\\
&& \times \Bigg(\int^{r}{Q(r')e^{\int^{r'}{P(r'')dr''}}dr'} + C_{A}\Bigg)\nb\\
&=& \sqrt{1 - \frac{2M}{r} -\frac{\Lambda_{g}}{3}r^2}\nb\\
&& \times \left(C_{A} - \frac{1}{6}\int^{r}{{\cal{D}}(r')dr'}\right),
%&& A = \sqrt{1 - \frac{2B}{r} -\frac{\Lambda_{g}}{3}r^2}\Bigg[C_A\nb\\
%&& -\frac{1}{6}\int\frac{\alpha_1r^9+\alpha_2r^6+\alpha_3r^3+\alpha_4r+\alpha_5}{\sqrt{r^{13}(r-2B-\frac{\Lambda_g}{3}r^3)^3}}\nb\\
%&& -\frac{4\pi G D_1^2r^5}{g_e^2\sqrt{r^{13}(r-2B-\frac{\Lambda_g}{3}r^3)^3}}dr\Bigg],
\eqn
where $C_A$ is an integration constant, and
\bqn
\lb{4.71ed}
{\cal{D}}(r) &\equiv & \frac{1} {\sqrt{r^{16}\left(1 - \frac{2M}{r} -\frac{\Lambda_{g}}{3}r^2\right)^{3}}} \Bigg({\alpha_{1}}{r^{9}}  + {\alpha_{2}}{r^{6}} \nb\\
&& + {\alpha_{3}}{r^{3}}  + {\alpha_{4}}r + {\alpha_{5}} - \frac{4\pi G D^{2}_{1}}{g_{e}^{2}}  r^{5}\Bigg).
\eqn

For the special case $\Lambda_g =0$, $A$ is given explicitly by % we have
\bqn
\lb{4.71ee}
A(r) &=& 1+\frac{D_1^2}{12g_e^2M^2}\left(\frac{M}{r}-1\right)+C_A\sqrt{1-\frac{2M}{r}}\nb\\
&& +\frac{g_0\zeta^2}{8}\Bigg[r^2+5Mr -  30M^2\Bigg(1 \nb\\
&& - \sqrt{1-\frac{2M}{r}}\ln\left(\sqrt{r}+\sqrt{r-2M}\right)\Bigg)\Bigg]\nb\\
&& +\frac{g_3}{10M^2r^3\zeta^2}\Big(M^3+M^2r+2Mr^2-2r^3\Big)\nb\\
&& -\frac{\alpha_4}{378M^6r^5}\Bigg(7M^5+5M^4r+4M^3r^2\nb\\
&& ~~~~~~~~~~~~~~~~~~~~ +4M^2r^3  +8Mr^4-8r^5\Bigg)\nb\\
&& -\frac{\alpha_5}{1386M^7r^6}\Bigg(21M^6+14M^5r+10M^4r^2\nb\\
&& +8M^3r^3 +8M^2r^4+16Mr^5-16r^6\Bigg). ~~~ %,\nb\\
%&& ~~~~~~~~~~~~~~~~~~~~~~~~~~~~~  \; (\Lambda_{g} = 0).
\eqn
%where we have put $B=G$.
%\newpage

\section{Charged Black Holes}

\renewcommand{\theequation}{5.\arabic{equation}} \setcounter{equation}{0}

The causal structure of spacetimes in the HL theory is different from that in GR, because of the breaking of the Lorentz symmetry. 
In particular, the dispersion relations of particles contain high-order momentum terms \cite{Horava,WM,WWa,CH},
\bq
\lb{5.1}
\omega_{k}^{2} = m^{2} + k^{2}\left(1 + \frac{k^{2}}{M_{A}^{2}}   + \frac{ k^{4}}{M_{B}^{4}}\right),
\eq
where $M_{A}$ and $M_{B}$ are the suppression energy scales of the fourth and sixth order derivative terms.
As an result,  the speeds of partilces $v_{p} \left(\equiv d\omega_k/dk\right)$  become unbounded in the UV, and their motions do not follow geodesics.   
This immediately makes all the  definitions of black holes given in GR invalid \cite{HE73,Tip77,Hay94,Wang}.  To provide a proper definition of black holes, 
anisotropic conformal boundaries \cite{HMT2} and  kinematics of particles  \cite{KM} have been studied   in the HL theory.
 In particular, in \cite{GLLSW} black holes and global structure of spacetimes were studied by defining  a horizon as the infinitely redshifted 2-dimensional 
 (closed) surface of massless test particles \cite{KK}.  Such a definition reduces to that given in GR when the dispersion relation is relativistic, where  
 $M_{A}, M_{B} \gg k$, as one can see from  Eq.(\ref{5.1}).  
 
 To study the black hole structure of the solutions presented in the last section, following \cite{GLLSW} let us consider
  a scalar field with a given dispersion relation $F(\zeta)$. In the geometrical optical  approximations,  
$\zeta$ is given by $\zeta = g_{ij}k^{i}k^{j}$, where $k_{i}$ denotes the 3-momentum of the corresponding  massless particle. With this  approximation, 
the trajectory of a test particle is given by
\bqn
\lb{eqq1}
S_{p} &\equiv& \int_{0}^{1}{{\cal{L}}_{p} d\tau} \nb\\
&=& \frac{1}{2} \int_{0}^{1}{ d\tau\Bigg\{\frac{c^{2}N^{2}}{e} \dot{t}^{2} + e \Big[F(\zeta) - 2 \zeta F'(\zeta)\Big]\Bigg\}}, ~~~~
\eqn
where $e$ is a one-dimensional einbein, and   $\zeta$ is now considered as a functional of $t, x^{i}, \dot{t}, \dot{x}^{i}$ and $e$,  given by the relation,
\bq
\lb{eqq2}
\zeta\;  [F'(\zeta)]^{2} = \frac{1}{e^{2}} g_{ij}\big(\dot{x}^{i} + N^{i} \dot{t}\big) \big(\dot{x}^{j} + N^{j}  \dot{t}\big),
\eq
with $\dot{t} \equiv dt/d\tau$, etc.   Considering Eq.(\ref{5.1}), we assume that  
\bq
\lb{F-function}
F(\zeta) = \zeta^{n}, \; (n = 1, 2, ...).
\eq
Then, Eq.(\ref{eqq2}) yields,
\bq
\lb{eqq4}
\zeta = \left(\frac{\dot{r} + N^{r}\dot{t}}{n e \sqrt{f}}\right)^{{2}/{(2n - 1)}} \equiv \left(\frac{{\cal{D}}}{e^{2}}\right)^{{1}/{(2n - 1)}},
\eq
where
\bq
\lb{5.2}
N^{r} = - e^{\mu - \nu},\;\;\; f = e^{-2\nu}.
\eq
Note that there is a sign difference between $N^{r}$ defined here and the one defined in Eq.(\ref{4.2}). This corresponds to the
coordinate transformation $t \rightarrow - t$, that is, in the static case, if $(N, N^r, \nu)$ is a solution of the HL theory, so is the
one  $(N, -N^r, \nu)$. Keeping this in mind, and inserting the above into Eq.(\ref{eqq1}), we find that, for  radially moving massless 
particles, 
${\cal{L}}_{p}$ is given by
\bq
\lb{eqq5}
{\cal{L}}_{p} = \frac{N^{2}}{2e} \dot{t}^{2} + \frac{1}{2} \big(1 - 2n\big)e^{1/(1-2n)} {\cal{D}}^{{n}/{(2n - 1)}}.
\eq
Then, the variations of   ${\cal{L}}_{p} = 0$ with respect to $e$ and $t$ yield, respectively, 
\bqn
\lb{eqq6}
N^{2}\dot{t}^{2} - e^{2(n-1)/(2n-1)} {\cal{D}}^{{n}/{(2n - 1)}} = 0,\\
\lb{eqq8}
N^{2}\dot{t} - e^{2(n-1)/(2n-1)}\frac{N^{r}}{\sqrt{f}} {\cal{D}}^{{1}/{[2(2n - 1)]}} = e E,
\eqn
where $E$ is an integration constant, representing the total energy of the test particle. 
eliminating $e$ from Eqs.(\ref{eqq6}) and (\ref{eqq8}) we find that
 \bq
 \lb{eqq16}
 X^{n} - p(r) X - q(r, E) = 0,
 \eq
 where
 \bqn
 \lb{eqq17}
 X &\equiv& \left(\frac{\sqrt{\cal{D}}}{\dot{t}}\right)^{1/(n-1)} = \left(\frac{\left|r' + N^{r}\right|}{n \sqrt{f}}\right)^{1/(n-1)} ,\nb\\
 p(r) &\equiv& \frac{N^{r}}{\sqrt{f}},\;\;\;
 q(r, E) \equiv E N^{1/(n-1)}, 
 \eqn
 with $r' \equiv \dot{r}/\dot{t} = dr/dt$. Once $X$ is found by solving Eq.(\ref{eqq17}), from it we obtain
 \bq
 \lb{5.3}
 t = t_{0} + \int{\frac{dr}{H(r, E)}},
 \eq
 where
 \bqn
 \lb{5.4}
 H(r, E) &=& \epsilon n e^{-\nu} X^{n-1} - N^{r}\nb\\
 &=& \left(\epsilon n -1\right) N^{r} + \epsilon n E \frac{N^{1/(n-1)}}{X}e^{-\nu},
 \eqn
 with $\epsilon = {\mbox{sign}}\; (\dot{r} + N^{r} \dot{t})$. In the last step of the above expressions, we used Eq.(\ref{eqq16})
 to replace $X^{n-1}$.   For detail, we refer readers to \cite{GLLSW}. 
 
   {\em A horizon is defined as a surface that is free of spacetime singularities and on which massless test particles are infinitely redshifted.} %, $|t| \rightarrow \infty$, as $r \rightarrow r_{H}$}.   
Note that the nature of  singularities  in the HL theory was studied in \cite{CW}, and was shown that they can be classified into two classes: the coordinate  singularities and spacetime singularities.
The coordinate singularities are the ones that can be removed by the general coordinate transformations (\ref{1.4}), while the spacetime singularities are  ones that cannot be
removed by  (\ref{1.4}). It should be noted that, although these definitions are the same as those  
given in GR, there are fundamental differences, because of the symmetry (\ref{1.4}) of the HL theory.  In \cite{CW}, some examples are given in which coordinate singularities in GR
become spacetime singularities in the HL theory. Spacetime singularities can be further divided into two   kinds:
 the curvature and non-curvature
ones. A curvature singularity is defined as the one in which at least one of the scalars of the symmetry (\ref{symmetry}) is singular. A non-curvature singularity is defined as the
one that does not have curvature singularity,  but some other physical quantities, such as tidal forces and/or distortions  experienced by a test particle,  become unbounded. 

Assuming that at a surface, say, $r_{H}$, $H$ defined by Eq.(\ref{5.4}) behaves as
\bq
\lb{5.5}
H(r, E) = H_{0}(r_{H}, E)\left(r - r_{H}\right)^{\delta} + ...,
\eq
as $r \rightarrow r_{H}$, where $H_0(r_{H}, E) \neq 0$.  
Then,
\bq
\lb{CT}
H'(r, E) \bigg|_{r = r_{H}} = \cases{ 0, & $ \delta > 1$, \cr
H_0(r_{H}, E), & $ \delta  = 1$, \cr
\pm \infty, & $0 <\delta < 1$.}
\eq
Now $t \to \infty$ as $r \rightarrow  r_{H}^{+}$ if and only if 
\bq
\lb{delta}
\delta \geq 1, % \; (n \ge 2),
\eq
for which we have
\bq
\lb{5.6}
\left. \frac{dH(r, E)}{dr} \right|_{r = r_{H}} =\;  {\mbox{  finite}}. 
 \eq
This provides the necessary  condition on  $f, N, N^r, E, n$ for the hypersurface $r_{H}$ to be a horizon, as defined above. %blow-up of $t$ at $r_{H}$.
It should be noted that $r_{H}$ usually depends on the energy $E$ of the test particles, as first noted in \cite{GLLSW}. 

 To study the black hole structure of the solutions presented in the last section, let us   consider the cases $N^{r} = 0$ and $N^{r} \not=0$
 separately.

 \subsection{$N^{r} \not= 0$}
 
 In this case, the solutions are given by Eqs.(\ref{4.12}) and (\ref{4.15}). Inserting them into Eq.(\ref{5.5}) and considering the Eq.(\ref{5.2}), we find that
 \bq
 \lb{5.7}
 H(r, E) = (1+n) e^{\mu} - \frac{nE}{X}.
 \eq
 Note that in writing the above expression, we had chosen $\epsilon  =-1$ \cite{GLLSW}. Thus, at $r = r_{H}$, we have
 \bq
 \lb{5.8}
%X_{H} \equiv  
X(r_{H}, E) = \frac{nE}{1+n} e^{-\mu(r_{H})}.
 \eq
 Inserting it into Eq.(\ref{eqq16}), we obtain 
 \bq
 \lb{5.9}
 \mu(r) - \mu_{(n, E)} = 0,
 \eq
 at $r = r_{H}$, where
 \bq
 \lb{5.10}
 \mu_{(n, E)} \equiv  \frac{1}{n}\ln\left[\frac{1}{n}\left(\frac{1+n}{nE}\right)^{n-1}\right].
 \eq
  On the other hand, from Eqs.(\ref{eqq16}) and (\ref{5.7}), we also have
  \bq
  \lb{5.11}
  H'(r_{H}, E) = \frac{1}{2}(1+n) e^{\mu_{(n, E)}} \mu'(r_{H}).
  \eq
Thus, the condition (\ref{5.6})  requires 
\bq
\lb{5.12}
\left. \mu'(r) \right|_{r = r_{H}} =\;  {\mbox{  finite}}. 
\eq

Inserting the general solution (\ref{4.15}) into Eq.(\ref{5.9}), one can find all the roots, $r = r_{H}$. Provided that the solutions have no spacetime singularities and the condition (\ref{5.12})
holds, the 2-sphere $r = r_{H}$ represents a horizon. As mentioned above, $r_{H}$  in general depends on $E$, that is, the radius of the horizon is observer-dependent. Such a dependence
is  the reflection of the fact that the HL theory breaks the Lorentz symmetry.  

To see the above explicitly, let us consider the case $A = $ Constant and $\Lambda = 0$, for which Eq.(\ref{4.15})  reduces to Reissner-Nordstr\"om solution found in GR but written in the
Painleve-Gullstrand coordinates \cite{GP}, %Arnowitt-Deser-Misner form (\ref{4.1}).
\bq
\lb{5.13}
\mu_{RN}(r) = \frac{1}{2}\ln\Bigg(\frac{r_{g}}{r}    - \frac{r_{Q}^{2}}{r^{2}}\Bigg),
\eq
where $r_{g} \equiv 2m,\; r_{Q}^{2} \equiv {4\pi G Q^{2}}/{g_{e}^{2}}$. Inserting it into Eq.(\ref{5.9}) we find that
\bq
\lb{5.14}
r_{H}^{\pm} = \frac{1}{2}e^{-2\mu_{(n, E)}}\Bigg(r_{g} \pm \sqrt{r^{2}_{g} - 4e^{2\mu_{(n, E)}} r^{2}_{Q}}\Bigg).
\eq
It can be shown that the solutions at $r_{H}^{\pm}$ is free of spacetime singularities and Eq.(\ref{5.12}) is satisfied.
Therefore, provided that $r^{2}_{g} - 4 r^{2}_{Q}e^{2\mu_{(n, E)}} \ge 0$, the solutions have two horizons at $r = r_{H}^{\pm}$. 
When $n = 1$, we have $\mu_{(n = 1, E)} = 0$, and the above expressions reduce exactly to those given in GR \cite{dInverno}.
However, when $ n > 1$, from Eq.(\ref{5.10}) we find that
\bq
\lb{5.15}
 \mu_{(n, E)} \simeq  - \frac{1+n}{n}\ln{E},
 \eq
 as $E \rightarrow \infty$. Then, Eq.(\ref{5.14}) show that $r_{H}^{\pm} \simeq 0$, that is, as long as the test particles have enough energy
 ($E \gg 1$), the horizons  can be as closed to the singularity at $r = 0$ as desired. A similar situation also happens to the Schwarzschild solution
 \cite{GLLSW}. Again, this is because of the violation of the Lorentz symmetry in the UV regime.

 When $A \not= $ Constant, the local and global structures of the corresponding spacetimes depend on the choice of $A(r)$ (as well as $\Lambda$).
 It is not difficult to see that the spacetimes have very rich structures, and some of them will quite similar to the Reissner-Nordstr\"om solution, and
 in general the radius $r_{H}$ will depend on the energy of the observers,  i.e., $r_{H} = r_{H}(E)$.

 \subsection{$N^{r} = 0$}
 
 When $N^{r} = 0$, we have $X = E^{1/n}$ and 
 \bqn
 \lb{5.16}
 H(r, E) &=& \epsilon n E^{(n-1)/n} e^{-\nu(r)}\nb\\
 &=& \epsilon n E^{(n-1)/n}\sqrt{1 - \frac{2M}{r} -\frac{\Lambda_{g}}{3}r^2}.
 \eqn
 To have a horizon, the necessary condition (\ref{delta}) requires that 
 \bq
 \lb{5.17}
 1 - \frac{2M}{r} -\frac{\Lambda_{g}}{3}r^2 = 0,
 \eq
  has at least two equal and positive roots. It is easy to show that this is
 not possible for any choice of $\Lambda_g$ and $M$. Therefore, this class of solutions does not have  black hole structures. The global structures
 of the spacetimes for various choices of the free parameters  $\Lambda_g$ and $M$ are given in \cite{GLLSW}, and we shall not repeat these
 analyses here.

\section{Conclusions}

\renewcommand{\theequation}{6.\arabic{equation}} \setcounter{equation}{0}

In this paper, we have studied electromagnetic  static spacetimes   in the Ho\v{r}ava and Melby-Thompson (HMT) setup \cite{HMT},  in which   $\lambda = 1$.
After writing down specifically the coupling of the theory with a massive vector field in Sec. III,  we have applied the general formulas to electric static spacetimes
with spherical symmetry in Sec. IV, and found  all the solutions  of the massless vector field under the gauge (\ref{gauge}).
In particular, when $N^{r} \not= 0$, the metric
coefficients are  given  by Eqs.(\ref{4.12})  and (\ref{4.15}) for any given gauge field $A$, subjected to the constraint (\ref{4.16}). The corresponding electromagnetic
 field $(A_{0}, A_{1}, 0, 0)$ is given by Eq.(\ref{4.13}). When $A = $ Constant, the solutions reduce to the  Reissner-Nordstr\"om
 one,  found in GR but written in the Painleve-Gullstrand coordinates \cite{GP}.

 In the diagonal case $N^{r} = 0$, the metric coefficients are  given  by Eqs.(\ref{NR})  and (\ref{4.71b}), while  the corresponding electromagnetic
 field $(A_{0}, A_{1}, 0, 0)$ and the gauge field $A$ are given, respectively, by Eqs.(\ref{4.71c}), (\ref{4.71d}) and (\ref{4.71ec}). When $\Lambda_{g} = 0$,
 the gauge field $A$ is explicitly given by Eq.(\ref{4.71ee}).
 
 In Sec. V, using the  geometrical optical approximations  \cite{GLLSW}, we have studied the existence of horizons  and shown explicitly why
 the definitions of black holes given in GR \cite{HE73,Tip77,Hay94,Wang} cannot be applied to the HL gravity, and how to generalize those definitions to
the HL theory, because of the breaking of the Lorentz-invariance, in which the conception  of light-cones is no longer valid. Applying the new definition
of horizons to our solutions presented in Sec. IV, we have found that some of the solutions with $N^{r} \not =0$ give rise to black holes, although
 the locations of the horizons  depend on the energies of the test particles. With sufficient high energy, the horizon can be as closed to the central singularity at origin as desired.
On the other hand, the solutions with $N^{r} = 0$ do not have black hole structures.  
It should be noted that one might argue that by the coordinate transformations, 
\bq
\lb{6.0}
d\tilde{r} = \frac{dr}{N^{r}(r)} + dt, 
\eq
one can bring the metric with $N^{r} \not= 0$  into the 
diagonal form,
\bq
\lb{6.1}
ds^{2} = - dt^{2} + e^{2\tilde{\nu}}d^{2}\tilde{r} + r^{2}d^{2}\Omega,
\eq
where $\tilde{\nu} = \nu(r) + \ln|N^{r}(r)|$. However, now  $r$ is time-dependent, $r = r(\tilde{r}, t)$ (and so is $\tilde\nu$), and then the trajectories of massless particles are no longer described by
Eqs.(\ref{eqq6}) and (\ref{eqq8}), so the discussions presented in Sec. V cannot be applied to the ``dynamical" case. % time-dependent case. 

 In addition,  the gauge field $A$ for the non-diagonal case (See Sec. IV.A)  is undetermined. While its physics  is not clear (even in more general case) \cite{HMT,WW}, the solar system tests
 generically require it  vanish \cite{GSW} (See also \cite{AP}). However, in the diagonal case it is uniquely determined by Eq.(\ref{4.71ec}). % the diagonal solutions (\ref{4.71b}),
 HMT showed that this class of solutions is consistent with observations only when the gauge field $A$  is considered as part of the lapse function in the IR, ${\cal{N}} = N - A$. In \cite{GSW} a different
 point of view was adopted, in which the gauge field as well as the Newtonian prepotential was considered as independent of the spacetime metric, although they are part of the gravitational field and
 interact with spacetime  through the field equations, roles quite similar to the Brans-Dicke scalar
 field in the Brans-Dicke theory of gravity \cite{BD}. This is seemingly supported by the results presented recently  in \cite{AdaS}. Moreover, some preliminary results of stability analysis show that this
 class of solutions might not be stable \cite{LW}. Clearly, to understand these solutions better, further investigations are highly demanded, including the possibilities of
 considering them as describing the spacetime outside of a charged star. The general junction conditions across the surface of a star have been worked out in \cite{GSW}, although internal solutions
 of charged stars have not been found, yet.

~\\{\bf Acknowledgements:}
The work of AW was supported in part by DOE  Grant, DE-FG02-10ER41692. KL was partially supported by NSFC  No. 11178018 and No. 11075224.

\section*{Appendix A:  $\left(F_{s}\right)_{ij}$ and $F_{(\varphi, n)}^{ij}$}
 \renewcommand{\theequation}{A.\arabic{equation}} \setcounter{equation}{0}

  $\left(F_{s}\right)_{ij}$ and $F_{(\varphi, n)}^{ij}$ appearing in Eq.(\ref{eq3a}) are given, respectively,
 \bqn
\lb{a.1}
(F_{0})_{ij}&=& -\frac12g_{ij},\nb\\
(F_{1})_{ij}&=&-\frac12g_{ij}R+R_{ij},\nb\\
(F_{2})_{ij} &=&-\frac12g_{ij}R^2+2RR_{ij}-2\nabla_{(i}\nabla_{j)}R\nb\\
                     & & +2g_{ij}\nabla^2R,\nb\\
(F_{3})_{ij}&=&-\frac12g_{ij}R_{mn}R^{mn}+2R_{ik}R^k_j-2\nabla^k\nabla_{(i}R_{j)k}\nb\\
                     && +\nabla^2R_{ij}+g_{ij}\nabla_m\nabla_nR^{mn},\nb\\
(F_{4})_{ij}&=&-\frac12g_{ij}R^3+3R^2R_{ij}-3\nabla_{(i}\nabla_{j)}R^2\nb\\
                    && +3g_{ij}\nabla^2R^2,\nb\\
(F_{5})_{ij}&=&-\frac12g_{ij}RR^{mn}R_{mn}+R_{ij}R^{mn}R_{mn}\nb\\
                    &&+ 2RR_{ki}R^k_j  -\nabla_{(i}\nabla_{j)}\left(R^{mn}R_{mn}\right)\nb\\
                    && - 2\nabla^n\nabla_{(i}RR_{j)n}  +g_{ij}\nabla^2\left(R^{mn}R_{mn}\right)\nb\\
                    &&  + \nabla^2\left(RR_{ij}\right)   +g_{ij}\nabla_m\nabla_n\left(RR^{mn}\right),\nb\\
(F_{6})_{ij}&=&-\frac12g_{ij}R^m_nR^n_pR^p_m+3R^{mn}R_{ni}R_{mj}\nb\\
                      && +\frac32\nabla^2\left(R_{in}R^n_j\right)   + \frac32g_{ij}\nabla_k\nabla_l\left(R^k_nR^{ln}\right)\nb\\
                      & & -3\nabla_k\nabla_{(i}\left(R_{j)n}R^{nk}\right),\nb\\
(F_{7})_{ij}&=&-\frac12g_{ij}(\nabla R)^2+ \left(\nabla_iR\right)\left(\nabla_jR\right) -2R_{ij}\nabla^2R\nb\\
&& +2\nabla_{(i}\nabla_{j)}\nabla^2R-2g_{ij}\nabla^4R,\nb\\
(F_{8})_{ij}&=& -\frac12g_{ij}\left(\nabla_p R_{mn}\right)\left(\nabla^p R^{mn}\right) -\nabla^4R_{ij}\nb\\
&&  + \left(\nabla_i R_{mn}\right)\left(\nabla_j R^{mn}\right) +2\left(\nabla_p R_{in}\right)\left(\nabla^p R^n_j\right)\nb\\
&&  +2\nabla^n\nabla_{(i}\nabla^2R_{j)n}+2\nabla_n\left(R^n_m\nabla_{(i}R^m_{j)}\right)\nb\\
&& -2\nabla_n\left(R_{m(j}\nabla_{i)}R^{mn}\right)-2\nabla_n\left(R_{m(i}\nabla^nR^m_{j)}\right)\nb\\
 && -g_{ij}\nabla^n\nabla^m\nabla^2R_{mn},
\eqn
and   \cite{WW}
\bqn
  \lb{eq3c}
F_{(\varphi, 1)}^{ij} &=& \frac{1}{2}\varphi\left\{\Big(2K + \nabla^{2}\varphi\Big) R^{ij}
- 2 \Big(2K^{j}_{k} + \nabla^{j} \nabla_{k}\varphi\Big) R^{ik} \right.\nb\\
& & ~~~~~ - 2 \Big(2K^{i}_{k} + \nabla^{i} \nabla_{k}\varphi\Big) R^{jk}\nb\\
& &~~~~~\left.
- \Big(2\Lambda_{g} - R\Big) \Big(2K^{ij} + \nabla^{i} \nabla^{j}\varphi\Big)\right\},\nb\\
F_{(\varphi, 2)}^{ij} &=& \frac{1}{2}\nabla_{k}\left\{\varphi{\cal{G}}^{ik}
\Big(\frac{2N^{j}}{N} + \nabla^{j}\varphi\Big) \right. \nb\\
& & \left.
+ \varphi{\cal{G}}^{jk}  \Big(\frac{2N^{i}}{N} + \nabla^{i}\varphi\Big)
-  \varphi{\cal{G}}^{ij}  \Big(\frac{2N^{k}}{N} + \nabla^{k}\varphi\Big)\right\}, \nb\\
F_{(\varphi, 3)}^{ij} &=& \frac{1}{2}\left\{2\nabla_{k} \nabla^{(i}f^{j) k}_{\varphi}
- \nabla^{2}f_{\varphi}^{ij}   - \left(\nabla_{k}\nabla_{l}f^{kl}_{\varphi}\right)g^{ij}\right\},\nb\\
\eqn
where
\bqn
\lb{eq3d}
f_{\varphi}^{ij} &=& \varphi\left\{\Big(2K^{ij} + \nabla^{i}\nabla^{j}\varphi\Big) %\right.\nb\\
%& & \left. ~~~~~~~~
- \frac{1}{2} \Big(2K + \nabla^{2}\varphi\Big)g^{ij}\right\}.\nb\\
\eqn

\section*{Appendix B:  $J^{t}, J_{i}, J_{\varphi}, J_{A}$ and $\tau_{ij}$ for a vector field}
 \renewcommand{\theequation}{B.\arabic{equation}} \setcounter{equation}{0}

When a vector field is the only source,  the  quantities $J^{t}, J_{i}, J_{\varphi}, J_{A}$ and $\tau_{ij}$    are given by
 \bqn
\lb{3.8}
J^{t} &=&\frac{1}{g_e^2}\Bigg[-\frac{1}{N^2}g^{ij}({\cal{F}}_{0i}-\hat{N}^k{\cal{F}}_{ki})({\cal{F}}_{0j}-\hat{N}^l{\cal{F}}_{jl})\nb\\
&&+\frac{1}{N}g^{ij}({\cal{F}}_{0i}-\hat{N}^k{\cal{F}}_{ki})(\nabla^l\varphi {\cal{F}}_{jl})\nb\\
&&+\frac{1}{N}g^{ij}({\cal{F}}_{0j}-\hat{N}^l{\cal{F}}_{jl})(\nabla^k\varphi {\cal{F}}_{ki})\nb\\
&&+\frac{m_e^2}{N}(A_0-\hat{N}^iA_i)(\nabla^j\varphi A_j)\nb\\
&& -\frac{m_e^2}{2N^2}(A_0-\hat{N}^iA_i)^2\Bigg], \nb\\
J_{i} &=& -\frac{1}{2g_e^2 N}g^{jk}\Bigg[({\cal{F}}_{0j}-\hat{N}^l{\cal{F}}_{jl}){\cal{F}}_{ki}\nb\\
&& +({\cal{F}}_{0k}-\hat{N}^l{\cal{F}}_{lk}){\cal{F}}_{ij}\Bigg]\nb\\
&&+\frac{m_e^2}{2g_e^2 N}(A_0-\hat{N}^kA_k)A_i,\nb\\
J_{\varphi} &=&-\frac{1}{\sqrt{g}N}\partial_t\Big[\sqrt{g}{\cal{K}} B_iB^i\Big]\nb\\
&&+\frac{m_e^2}{2g_e^2 N}\nabla^l\Big[(A_0-\hat{N}^iA_i)A_l\Big]\nb\\
&&-\frac{1}{N}\nabla_j\Big[{\cal{K}}B_iB^i(N\nabla^j\varphi-N^j)\Big] \nb\\
&&+ \frac{1}{2g_e^2 N}g^{ij}\nabla^l\Big[({\cal{F}}_{0i}-\hat{N}^k{\cal{F}}_{ki}){\cal{F}}_{jl}\nb\\
&&({\cal{F}}_{0j}-\hat{N}^k{\cal{F}}_{jk}){\cal{F}}_{li}\Big],\nb\\
J_{A} &=& 2{\cal{K}} B_iB^i,\nb\\
  \tau^{mn} &=&g^{mn}{\cal{L}}_{EM}\nb\\
&&-\frac{1}{2g_e^2 N^2}\Bigg[({\cal{F}}_0^{~m}-\hat{N}_k{\cal{F}}^{km})({\cal{F}}_0^{~n}-\hat{N}_l{\cal{F}}^{nl})\nb\\
&&+({\cal{F}}_0^{~n}-\hat{N}_k{\cal{F}}^{kn})({\cal{F}}_0^{~m}-\hat{N}_l{\cal{F}}^{ml})\nb\\
&&+N({\cal{F}}_{0j}-\hat{N}^l{\cal{F}}_{jl})({\cal{F}}^{mj}\nabla^n\varphi+{\cal{F}}^{nj}\nabla^m\varphi )\nb\\
&&+N({\cal{F}}_{0i}-\hat{N}^l{\cal{F}}_{li})({\cal{F}}^{im}\nabla^n\varphi +{\cal{F}}^{in}\nabla^m\varphi)\Bigg]\nb\\
&&-\frac{m_e^2}{2g_e^2 N}(A_0-\hat{N}^iA_i)(A^m\nabla^n\varphi+A^n\nabla^m\varphi)\nb\\
&&- \frac{1}{2g_e^2}\Bigg\{a_1(B^mB^n-B_iB^ig^{mn})\nb\\
&& -a'_0A^mA^n - a'_1 A^mA^nB_jB^j\nb\\
&&- a'_2 A^mA^n(B_jB^j)^2\nb\\
&&+2a_2B_lB^l(B^mB^n-B_iB^ig^{mn})\nb\\
&&- a'_3A^mA^n(B_jB^j)^3+3a_3(B_lB^l)^2(B^mB^n\nb\\
&&-B_iB^ig^{mn})- a'_4 A^mA^n\left(\nabla_l B_j\right)\nabla^l B^j\nb\\
&&-a_4\Big[\left(\nabla^mB_j\right)\nabla^nB^j+\left(\nabla_iB^m\right)\nabla^iB^n\Big]\nb\\
&&+\Big[\nabla_i(a_4\nabla^iB^j)\Big]B_jg^{mn}-2\Big[\nabla_i(a_4\nabla^iB^{(n})\Big]B^{m)}\nb\\
&&+\nabla_i\Bigg[a_4\Big(B^{(m}\nabla^iB^{n)}+B^{(m}\nabla^{n)}B^i\nb\\
&& -B^i\nabla^{(m}B^{n)}\Big)\Bigg] - a'_5 A^mA^nB_kB^k\left(\nabla_l B_j\right)\nabla^l B^j\nb\\
&&+a_5\left(\nabla_l B_k\right)\left(\nabla^l B^k\right)(B^mB^n-B_iB^ig^{mn})\nb\\
&&-a_5B_kB^k\Big[\left(\nabla^mB_j\right)\nabla^nB^j+\left(\nabla_jB^m\right)\nabla^jB^n\Big]\nb\\
&&+\Big[\nabla_i(a_5B_kB^k\nabla^iB^j)\Big]B_jg^{mn}\nb\\
&&-2\Big[\nabla_i(a_5B_kB^k\nabla^iB^{(n})\Big]B^{m)}\nb\\
&&+\nabla_i\Bigg[a_5B_kB^k\Big(B^{(m}\nabla^iB^{n)}+B^{(m}\nabla^{n)}B^i\nb\\
&&-B^i\nabla^{(m}B^{n)}\Big)\Bigg]\nb\\
&& - a'_6 A^mA^n\left(\nabla_l B_j\right)\left(\nabla^lB^k\right)\nabla^jB_k\nb\\
&&+a_6\left(\nabla_iB^j\right)\left(\nabla^iB^{(m}\right)\nabla_jB^{n)}\nb\\
&&  -\frac{1}{2}a_6\Big[\left(\nabla^nB^j\right)\left(\nabla^mB^k\right)\nabla_j B_k\nb\\
&&+\left(\nabla^mB^j\right)\left(\nabla^nB^k\right)\nabla_j B_k\Big]\nb\\
&& +\frac{1}{2}B^j g^{mn}\nabla_i(a_6\alpha^i_j)\nb\\
&&-\frac{1}{2}\nabla_i\Bigg[a_6\Big(B^i\alpha^{(mn)}+B^{(m}\alpha^{in)}-B^{(n}\alpha^{m)i}\Big)\Bigg]\nb\\
&&- a'_7 A^mA^n(\nabla_l \nabla_j B_k)( \nabla^l\nabla^jB^k)\nb\\
&&- a_7\Big[(\nabla^m\nabla_j B_k)(\nabla^n\nabla^j B^k)\nb\\
&&+ (\nabla_i\nabla^m B_k)(\nabla^i\nabla^n B^k)\nb\\
&& -(\nabla_i\nabla_j B^n)(\nabla^i\nabla^j B^m)\Big]\nb\\
&&-B^lg^{mn}\nabla_j\nabla_i(a_7\nabla^i\nabla^jB_l)\nb\\
&&+\nabla_k\Big(\beta^{k(mn)}-\beta^{(mn)k}-\beta^{(mkn)}\Big)\Bigg\}\nb\\
&&-\frac{2}{N}\Bigg[{\cal{K}}'A^mA^nB_jB^j(A-{\cal{A}})\nb\\
&&-{\cal{K}}\Big(B^mB^n-B_iB^ig^{mn}\Big)(A-{\cal{A}})\nb\\
&&-\frac{N}{2}{\cal{K}}B_iB^i(\nabla^m\varphi)(\nabla^n\varphi)\Bigg],
\eqn
where
\bqn
\alpha^i_j &\equiv& (\nabla^iB^k)(\nabla_j B_k)+(\nabla^iB^k)(\nabla_k B_j)\nb\\
&&+(\nabla_kB^i)(\nabla^k B_j), \nb\\
\beta^{kil} &\equiv& -\nabla_j\Big(a_7\nabla^j\nabla^iB^k\Big)B^l %\nb\\ &&
-a_7(\nabla^i\nabla^lB_j)\nabla^kB^j\nb\\
&& +a_7(\nabla^i\nabla^jB^k)\nabla_jB^l.
\eqn

%%%%%%%%%%%%%%%%%%%%%%%%%%%%%%%%%%%%%%%%%%%%%%%%%%%%%%

\end{document}